# Analytic Solution of Multi-Dimensional Schrödinger in Hot and Dense QCD Media Using SUSYQM Method

M. Abu-Shady[1*] and A. N. Ikot[2**]

[1]Department of Applied Mathematics, Faculty of Science, Menoufia University, Egypt

[2]Department of Physics, University of Port Harcourt, Nigeria

## Abstract

The N-radial Schrödinger equation is analytically solved by using SUSYQM method, in which the heavy quarkonia potential is introduced at finite temperature and baryon chemical potential. The energy eigenvalue is calculated in the N-dimensional space.  The obtained results show that the binding energy strongly decreases with increasing temperature and is slightly sensitive for changing baryon chemical potential up to 0.6 GeV at higher values of temperatures. We employed the nonperturbative corrections to the leading-order of the Debye mass at finite baryon chemical potential. We found that the binding energy is more dissociates when the nonperturbative corrections are included with the leading-order term of  Debye mass in both hot and dense media.  A comparison is discussed with other works such as the lattice parameterized of Debye mas. Thus, the present potential with the SUSYQM method provides satisfying results for the description of the dissociation of binding energy for heavy quarkonia in hot and dense media.

# 1. Introduction

The Supersymmetry quantum mechanics (SUSYQM) and shape invariance in physics [1-3] have been used extensively in finding exact and approximate solutions of the Schrödinger-like equations with solvable potential models [4-10]. It is well-known that SUSYQM allows one to determine the eigenvalues and eigenfunctions analytically for solvable potentials model using algebraic operator formulation without solving the Schrödinger-like differential equation by the standard series method. Many authors have in recent times solved the Schrodinger-like equation with physically motivated potential models [11-15]. The SUSYQM was first introduced by Witten [1] for the first time as a simplest supersymmetric model in quantum field theory. Different analytical techniques have been employed by many researchers for solving Schrödinger equation with physically motivated potentials. These methods include Nikiforov-Uvraov (NU) method [16-18], asymptotic iteration method (AIM)[19-20], shape invariance [1-3], SUSYQM [1-4 ], factorization method [21], and others [22-23].

The studying of quarkonia at finite temperature is an essential tool for understanding the status of the matter (confined/confined) formed in the heavy ion collisions such in Refs. [24-27]. Many attempts have introduced to calculate the dissociation of binding energy of quarkonium in the quark-gluon plasma by applying lattice calculations [28-31] or non-relativistic quark models such as Schrödinger equation [23-37]. Thus, the solution of the Schrödinger equation at finite temperature is a good tool for this target. The quarkonium properties have been studied by modifying both the Coulombic and string terms of the heavy quark potential using the perturbative hard thermal loop (HTL) dielectric permittivity in both the isotropic and anisotropic media in the static [ 38-46].

So far, most of the attempts concentrate on the study of the heavy bound-state properties at vanishing baryonic chemical potential of the thermal medium based on leading-order of the Debye mass such as in Refs. [38-43] using numerical methods to solve the Schrödinger equation. In Refs. [47, 48] have investigated the quarkonium dissociation at finite temperature and chemical potential using leading-order of Debye mass in their calculations.

The aims of this work: First, the analytic solution of N-Schrödinger equation using SUSYQM. So far no attempts for solving N-Schrödinger equation in hot and dense media using SUSYQM to the best of our knowledge has been reported. Second, applying present results to calculate the binding energy which depends on different types of Debye mass in hot and dense media which are not considered in other works.

The paper is organized as follows: In Sec. 2, the multi-dimensional Schrödinger equation is introduced with heavy quark potential in hot and dense media. Moreover, the solution of the N-dimensional Schrödinger equation is given by using the SUSYQM method. In Sec. 3, the binding energy is studied with different forms of Debye mass in the three-dimensional space. In Sec. 4, the summary and conclusion are presented.

## 2. Approximate Solutions of the N-dimensional Radial Schrödinger Equation with Quarkonium potential

### 2.1 Energy eigenvalue

The N-dimensional radial Schrödinger equation for two particles interacting with asymmetric potential $V(r)$ takes the form [49, 50],

$$\left\{\frac{d^2}{dr^2}+\frac{N-1}{r}\frac{d}{dr}+2\mu\left(E_{nl}-V(r)-\frac{l(l+N-2)}{2\mu r^2}\right)\right\}\psi(r)=0 \qquad (1)$$

where $\mu, l, N$ and $E$ are the reduced mass of the quarkonium, the orbital momentum quantum number, the dimensionality number, and the energy eigenvalue, respectively. Defining an ansatz for the wave function as $\psi(r)=r^{\left(\frac{1-N}{2}\right)}R(r)$, then Eq. (1) becomes,

$$\left\{\frac{d^2}{dr^2}+2\mu\left[E_{nl}-V(r)-\frac{\left[\left(l+\frac{N-2}{2}\right)^2-\frac{1}{4}\right]}{2\mu r^2}\right]\right\}R(r)=0. \qquad (2)$$

In the present work, we take the heavy-quark potential in the hot and dense QCD media as in Ref. [47] as follows

$$V(r)=\left(\frac{2\alpha}{m_D^2}-\alpha\right)\frac{e^{-m_D r}}{r}-\frac{2\sigma}{m_D^2 r}+\frac{2\sigma}{m_D}-\alpha m_D, \qquad (3)$$

where, $\alpha$ and $\sigma$ are parameters determined later. $M_D(T, u_b)$ is Debye mass depends on temperature and baryon chemical potential.

By substituting by Eq. (3) into Eq. (2), we obtain

$$\frac{d^2 R(r)}{dr^2}+\left[2\mu E_{nl}-2\mu\left(\left(\frac{2\alpha}{m_D^2}-\alpha\right)\frac{e^{-m_D r}}{r}-\frac{2\sigma}{m_D^2 r}+\frac{2\sigma}{m_D}-\alpha m_D\right)-\frac{\left(l+\frac{N-2}{2}\right)^2-\frac{1}{4}}{r^2}\right]R(r)=0, \qquad (4)$$

To avoid the difficulty in the term $r^{-1}$ by taking the following approximation of the form [4],

$$\frac{1}{r} \approx \frac{m_D}{\left(1-e^{-m_D r}\right)}, \tag{5}$$

On substituting Eq. (5) into Eq.(4), we obtain a second-order Schrödinger equation as follows,

$$-\frac{d^2 R_{nl}(r)}{dr^2} + V_{eff}(r) R_{nl}(r) = \tilde{E}_{nl} R_{nl}(r) \tag{6}$$

where,

$$V_{eff}(r) = \frac{A e^{-2m_D r} + B e^{-m_D r} + C}{\left(1-e^{-m_D r}\right)^2}, \tag{7}$$

with

$$A = -2\mu m_D \left(\frac{2\sigma}{m_D^2} - \alpha\right) \tag{8}$$

$$B = 2\mu m_D \left(\frac{2\sigma}{m_D^2} - \alpha\right) + \frac{4\mu\sigma}{m_D} \tag{9}$$

$$C = m_D^2 \left[\left(l + \frac{N-2}{2}\right)^2 - \frac{1}{4}\right] - \frac{4\mu\sigma}{m_D}. \tag{10}$$

The effective energy from equation (6) is given as

$$\tilde{E}_{nl} = \left(\frac{2\mu E_{nl}}{\hbar^2} + \frac{2\mu\alpha m_D}{\hbar^2} - \frac{4\mu\sigma}{\hbar^2 m_D}\right). \tag{11}$$

The ground-state wave function $R_{0,l}(r)$ in supersymmetric quantum mechanics is defined as in Refs. [1-4]

$$R_{0,l}(r) = \exp(-\int W(r) dr), \tag{12}$$

Here the integrand $W(r)$ in equation (12) is called the superpotential and the Hamiltonian of the system is composed of the raising and lowering operators defined as in Refs. [1-4]

$$H_- = \hat{A}^+\hat{A} = -\frac{d^2}{dr^2} + V_-(r), \tag{13}$$

$$H_+ = \hat{A}\hat{A}^+ = -\frac{d^2}{dr^2} + V_+(r), \tag{14}$$

where, the lowering and raising operator are given as,

$$\hat{A} = \frac{d}{dr} - W(r), \tag{16}$$

$$\hat{A}^+ = -\frac{d}{dr} - W(r), \tag{17}$$

the partner Hamiltonian is obtained as follows

$$V_\pm(r) = W^2(r) \mp W'(r) \tag{18}$$

The associated Riccati equation in SUSYQM is of the form

$$W^2(r) \mp W'(r) = V_{eff}(r) - \tilde{E}_{0,J}, \tag{19}$$

Now with the effective potential of Eq. (7), we proposed the superpotential in the form,

$$W(r) = \frac{Pe^{-m_D r}}{\left(1 - e^{-m_D r}\right)} + Q \tag{20}$$

Putting Eq. (7) and Eq. (18) into Eq. (17), we obtain the following sets of expressions,

$$P = \frac{-m_D \pm \sqrt{m_D^2 + 4(A + B + C)}}{2} \tag{21}$$

$$Q = \frac{P^2 + C - A}{2P} \tag{22}$$

$$\tilde{E}_{0,l} = C - Q^2. \tag{23}$$

The two partner Hamiltonians can be obtained as [1-4]

$$V_-(r) = W^2(r) - \frac{dW}{dr}$$

$$= \frac{P(P-m_D)e^{-2m_D r}}{(1-e^{-m_D r})^2} + \frac{2P\left(\frac{P^2+C-A}{2P}\right)e^{-m_D r}}{(1-e^{-m_D r})} + \left(\frac{P^2+C-A}{2P}\right)^2 \tag{24}$$

$$V_+(r) = W^2(r) + \frac{dW}{dr}$$

$$= \frac{P(P+m_D)e^{-2m_D r}}{(1-e^{-m_D r})^2} + \frac{2P\left(\frac{P^2+C-A}{2P}\right)e^{-m_D r}}{(1-e^{-m_D r})} + \left(\frac{P^2+C-A}{2P}\right)^2 \tag{25}$$

The two potentials partner in equations (24) and (25) satisfy the relationship

$$V_+(r,\rho_0) = V_-(r,\rho_1) + R(\rho_1) \tag{26}$$

with $\rho_0 = P$ and $\rho_i$ is a function of $\rho_0$, i.e $\rho_1 = f(\rho_0) = \rho_0 - m_D$. Therefore, $\rho_n = f(\rho_0) = \rho_0 - nm_D$. Thus, we can see that the shape invariance holds via a mapping of the form $P \to P - m_D$. From Eq. (24), we get

$$R(\rho_1) = \left(\frac{(\rho_0)^2 + C - A}{2\rho_0}\right)^2 - \left(\frac{(\rho_1)^2 + C - A}{2\rho_1}\right)^2,$$

$$R(a_2) = \left(\frac{(\rho_1)^2 + C - A}{2\rho_1}\right)^2 - \left(\frac{(\rho_2)^2 + C - A}{2\rho_2}\right)^2, \tag{27}$$

$$\vdots$$

$$R(a_n) = \left(\frac{(\rho_{n-1})^2 + C - A}{2\rho_{n-1}}\right)^2 - \left(\frac{(\rho_n)^2 + C - A}{2\rho_n}\right)^2,$$

The total energy eigenvalues for the system is obtained as follows

$$\tilde{E}_{nl} = \tilde{E}_{nl}^- + \tilde{E}_{0,l} \tag{28}$$

where,

$$\tilde{E}_{nl}^- = \sum_{k=1}^{n} R(a_k) = \left(\frac{(\rho_0)^2 + C - A}{2\rho_0}\right)^2 - \left(\frac{(\rho_n)^2 + C - A}{2\rho_n}\right)^2. \tag{29}$$

Using Eqs.(20), (21) and (27) in Eq. (26), we obtain the energy spectrum for the quarkonium potential in the N-dimensional space as,

$$E_{nl} = -\frac{1}{8\mu m_D^2}\left(\frac{m_D^2\left[\left(l+\frac{N-2}{2}\right)^2 - \frac{1}{4}\right] - \frac{4\mu\sigma}{m_D} + 2\mu m_D\left(\frac{2\sigma}{m_D^2} - \alpha\right)}{\left(n+l+\frac{N-1}{2}\right)} + m_D^2\left(n+l+\frac{N-1}{2}\right)\right)^2$$

$$-\alpha m_D + \frac{m_D^2}{2\mu}\left[\left(l+\frac{N-2}{2}\right)^2 - \frac{1}{4}\right].$$

(30)

## 3. Discussions of Results

In the previous section, we obtained the solution of the N-Schrödinger equation for heavy-quarkonium potential which is given in Eq. (3). We obtained the analytic solution of Eq. (2) that is given in Eq. (30) in the N-space, in which the eigenvalue of energy is calculated as a function of temperature and baryon chemical potential. The analytic solution of the N-Schrödinger equation for heavy-quarkonium potential at finite temperature and chemical potential using SUSYQM are not considered in other works as we mentioned in the introduction of the paper.

At N = 3, we apply the present results for studying the behavior of binding energy in hot and dense media in the quark-gluon plasma. We employ here three forms of the Debye mass, the leading order term in QCD coupling $m_D^{Lo}$, nonpertutbatve corrections to leading order term of Debye mass, and the lattice parameterized form as in Refs. [38, 51, 52]. The forms are followed

$$m_D^{Lo} = g(T)T\sqrt{\frac{N_c}{3} + \frac{N_f}{6} + \frac{N_f}{2\pi^2}\left(\frac{u_b}{3T}\right)^2}, \qquad (31)$$

$$m_D^{NL} = m_D^{LO} + g(T)\sqrt{\frac{N_c T m_D^{LO}}{2\pi\pi}}\sqrt{\left(-\frac{1}{2} - \frac{\sqrt{\nu^2 + mg^2}}{2m_D^{LO}} + \ln\frac{2m_D^{Lo} + \sqrt{\nu^2 + mg^2}}{\sqrt{\nu^2 + mg^2}}\right)}, \qquad (32)$$

where

$$\nu = \frac{g^2 N_c T m_D^{LO}}{8\pi}\left(\ln\frac{4\pi}{g^2} + 1\right), mg = \frac{15g(T)^4 T^2}{16\pi^2}, \qquad (33)$$

$$m_D^{Lattice} = 1.4 m_D^{LO}, \qquad (34)$$

where $N_f$ is the number of flavors, $N_c$ is the number of colors, T is temperature and $u_b$ is the baryon chemical potential. $g(T)$ is the two loop expression for the QCD coupling constant at finite temperature and finite baryon chemical [47, 53]. In the present work, we define binding energy $E_B$ =$V(r \to \infty)$-$E_{nl}$ based on in Ref. [54]

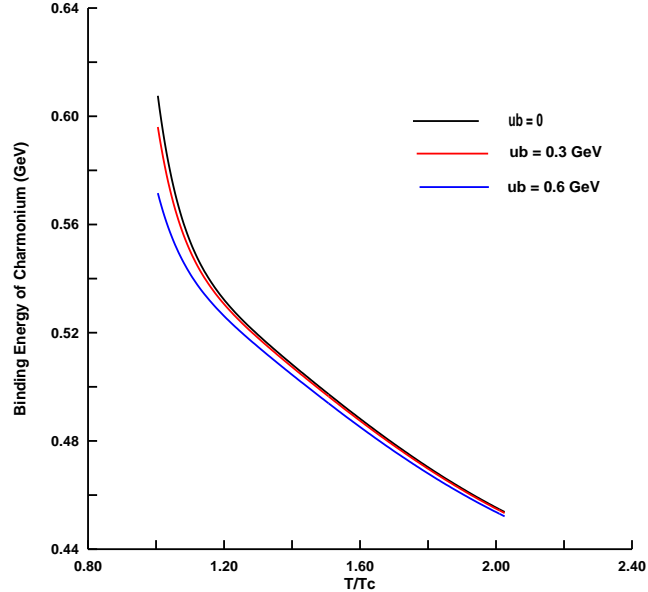

**In Fig. (1). The binding energy of 1S state of Charmonium is plotted as a function of temperature ratio for the leading-order term of Debye mass at different values of baryon chemical potential.**

In Fig. (1), the binding energy against the temperature is plotted. It is observed that the binding energy decreases with increasing temperature. The temperature dependence on the binding energies shows a qualitative agreement with the similar variations shown as Refs. [38, 39, 40]. Fig. (1) further shows that the baryon chemical potential is slightly sensitive at higher temperature above 1.4 $T_c$. We note that the binding energy decreases with increasing baryon chemical potential around $T_c$. This finding is in an agreement with Refs. [47].

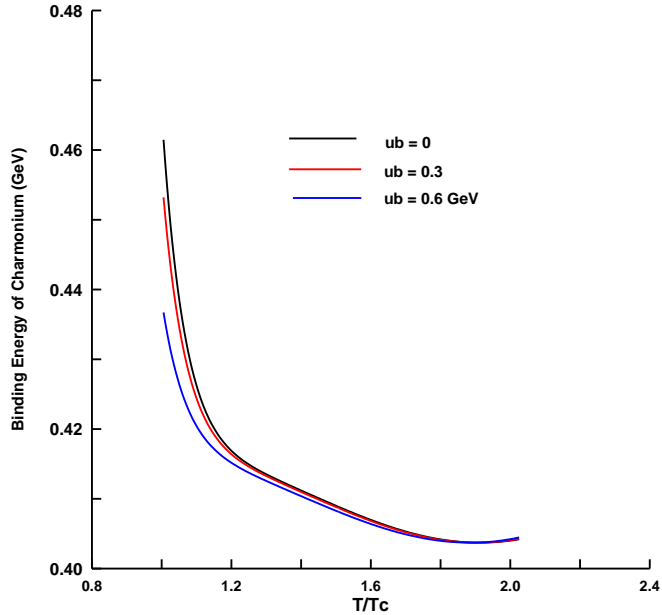

**In Fig. (2). The binding energy of 1S state of Charmonium is plotted as a function of temperature ratio for lattice QCD Debye mass at different values of baryon chemical potential.**

In Fig. (2), the binding energy is plotted based on the lattice QCD Debye mass. We note that binding energy decreases with increasing temperature. This qualitative agreement with the behavior of binding energy is defined through leading-order of Debye mass. We note that the binding energy has smaller values in comparison with the results of Fig. (1). This conclusion is in an agreement with that reported Ref. [38].

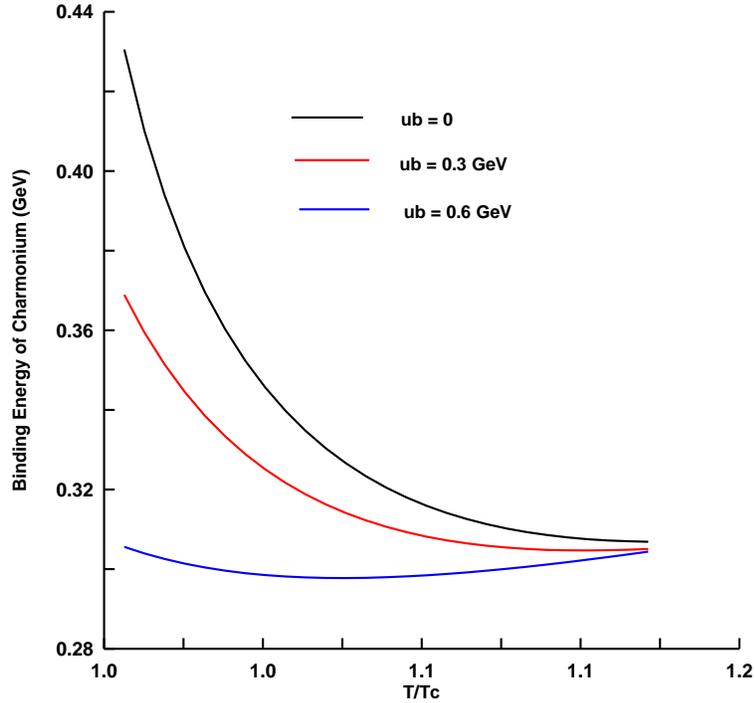

**In Fig. (3). The binding energy of 1S state of Charmonium is plotted as a function of temperature ratio for the next leading-order of Debye mass at different values of baryon chemical potential**

In Fig. (3), the binding energy is plotted using the next-leading-order term of Debye mass that is defined in Eq. (32). In previous works [44, 46, 47, 48], they concentrated on leading-order of Debye mass. In Ref. [38], the authors studied the effect of the leading-order of Debye mass at finite temperatures only. In the present work, we extend the previous works to finite baryon chemical potential. In Fig. (3), we note that the binding energy decreases with increasing temperature. In addition, we show that the binding energy strongly shifts to lower values by increasing baryon chemical potential closes 1.1 $T_c$. In addition, we note that the next leading-order of Debye mass strongly acts on reducing binding energy. This finding is found in Ref. [38].

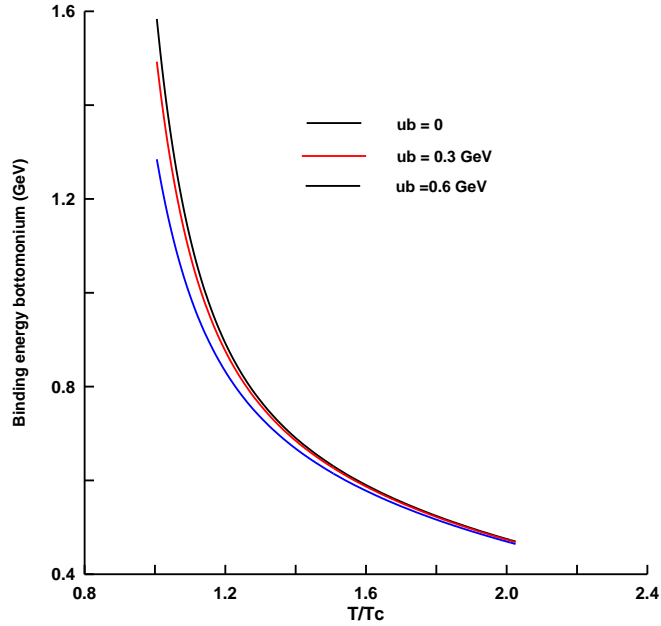

**In Fig. (4). The binding energy of 1S state of bottomonium is plotted as a function of temperature ratio for the leading-order term of Debye mass at different values of baryon chemical potential**

A similar situation is reported for the bottomonium. In Fig. (4), the binding energy is plotted as a function of temperature using Eq. (32). 1S state of bottomonium decreases with increasing temperature. In addition, the binding energy decreases with increasing baryon chemical potential. The strong effect of baryon chemical potential appears around critical temperature $T_c$ up to $1.2T_c$. In the present work, $T_c$=198 MeV as in Ref. [38] for $N_f$ = 2 and $N_c$ = 3, and $\sigma = 0.184$ GeV, $\alpha = \dfrac{g^2(T)}{4\pi}$ as Refs. [38, 47, 53]. In higher temperatures, the effect of chemical temperature is a little sensitive. The reason for this is that the chemical potential reflects only the baryon density in QGP but does not describe the color charge density [52]. In Ref. [47], the authors calculated the binding energy using the Schrödinger equation at finite temperature and chemical potential. They found the binding energy decreases with increasing temperature and chemical potential, however, their calculations carried out in low baryon chemical potential and leading-order term of Debye mass. We conclude that the behavior of the binding energy is an agreement with the results of Ref. [47]. Also, the present of 1S state of bottomonium is an agreement with Refs. [54, 55], in which the calculations are achieved in hot medium only.

In comparison with the Debye mass that deduced from lattice QCD, we have the similar behavior in comparison with binding energy using leading-order of Debye mass, however, we note that the binding energy of Fig. (5) provides smaller values in comparison with the binding energy of Fig. (4). Therefore, the binding energy rapidly dissociates in hot and dense media. In Fig. (6), the binding energy is plotted using nonperturbative corrections that defined in Eq. (32). We note that the binding energy decreases with increasing finite temperature and chemical potential and we note the binding energy gives smaller values in comparison with binding energy that deduced from the leading-order term of Debye mass in Fig. (4).

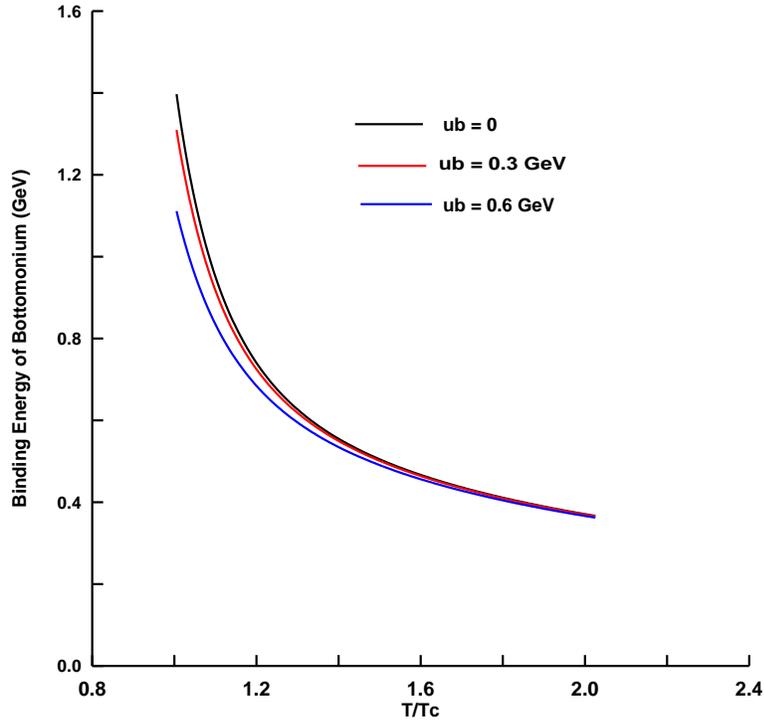

**In Fig. (5). The binding energy of 1S state of bottomonium is plotted as a function of temperature ratio for the lattice QCD Debye mass at different values of baryon chemical potential.**

This clearly appears in Fig. (7) where the binding energy has smaller values in comparison with the binding energy depends on leading-order term. Therefore, the binding energy rapidly dissociates in hot and dense media depends on the next leading-order term. One of advantages of present work that most of previous works did not considered the next leading-order term of Debye mass at finite chemical potential in their calculations. In Fig. (8), we display the binding energy for 1S state of bottomonium as a function of temperature and chemical potential depends on the leading-order of Debye mass. We note that the binding energy strongly depends on the temperature and a little sensitive on the baryon chemical potential.

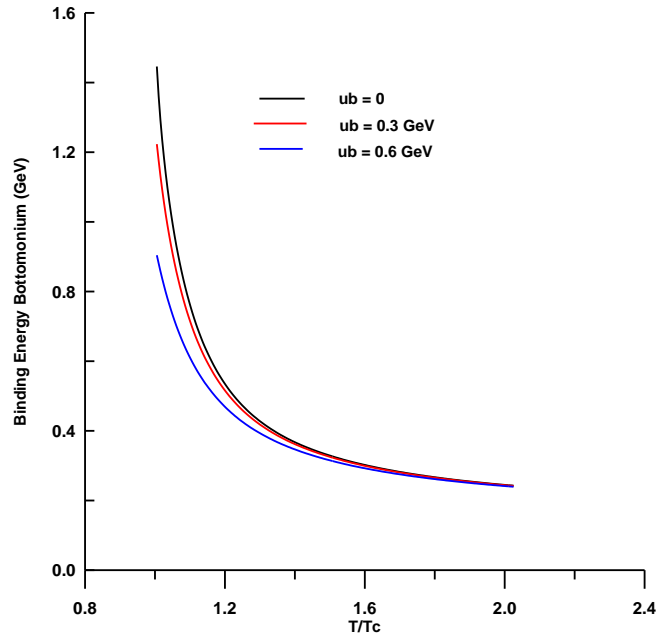

**In Fig. (6). The binding energy of 1S state of bottomonium is plotted as a function of temperature ratio for the next leading-order of Debye mass at different values of baryon chemical potential**

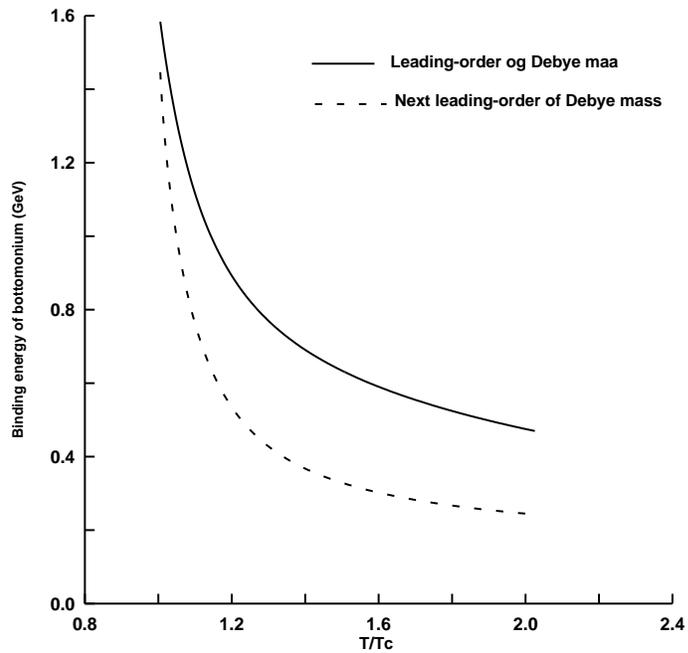

**In Fig. (7). The binding energy of 1S state of bottomonium is plotted as a function of temperature ratio for the leading-order of Debye mass and the next leading-order of Debye mass at $u_b = 0$**

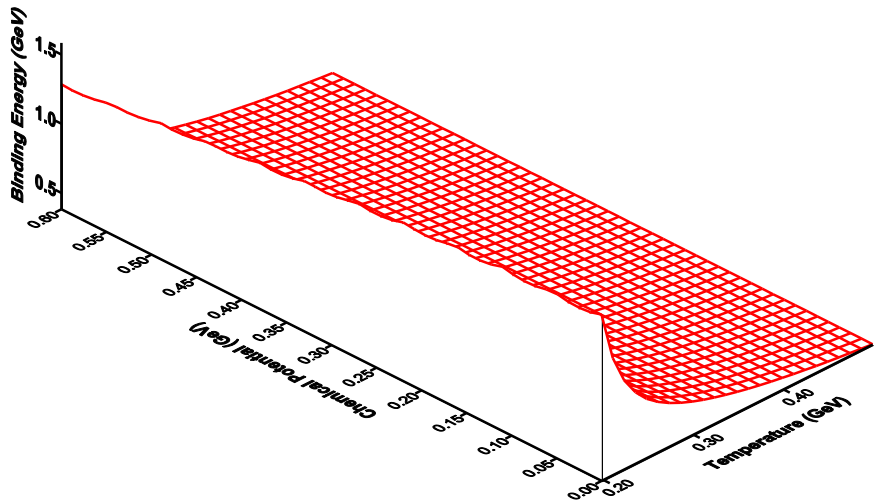

**In Fig. (8). The binding energy of 1S state of bottomonium is plotted as a function of temperature and baryon chemical potential for leading-order term of Debye mass.**

## 4. Summary and Conclusions

In the present work, the N-radial Schrödinger equation is analytically solved by using SUSYQM method. The eigenvalues of energy are obtained in the N-dimensional space. The eigenvalues of energy can be obtained in the lower dimensions. The solution of Schrödinger is analytically solved in hot and dense media. So far no attempts to solve quarkonium potential using SUSYQM was reported Refs. [4-10, 38-55].

We have applied the present results to calculate the binding energy and we have discussed the behavior of binding energy at finite temperature (hot medium) and finite chemical potential (dense medium). We have found that the binding energy strongly decreases with increasing temperature and a little sensitive to baryon chemical potential at higher values of temperatures. We have discussed the effect of kind of Debye mass on the binding energy. We have found the nonpertutbatve corrections to leading-order of Debye mass leads to the binding energy is more dissociates in hot and dense media. Also, the binding energy depends on the lattice QCD of Debye mass has smaller values than the leading-order of Debye mass. Generally, we note also that the qualitative agreement for the binding energy of charmoniun and bottomonium for different forms of Debye mass. A comparison with other works is discussed; we have found that the present results are in a good agreement with Refs. [38, 47] at finite temperature. One of advantages of present work that we have considered nonpertutbatve corrections to leading-order of Debye mass at finite chemical potential. Thus, the present work provides good description for quarkonium in hot and dense media to analysis the baryon rich quark gluon plasma which is expected at FAIR energies will be created.